\documentclass[twocolumn,superscriptaddress,pra,aps]{revtex4-1}
\usepackage{amssymb}
\usepackage{amsmath}
\usepackage{graphicx}
\usepackage{epsfig}

\begin{document}

\title{Symmetry protected topological phases characterized by isolated
exceptional points}
\author{S. Lin}
\author{L. Jin}
\email{jinliang@nankai.edu.cn}
\author{Z. Song}
\email{songtc@nankai.edu.cn}
\affiliation{School of Physics, Nankai University, Tianjin 300071, China}
\begin{abstract}
Exceptional point (EP) associated with eigenstates coalescence in
non-Hermitian systems has many exotic features. The EPs are generally
sensitive to system parameters, here we report symmetry protected isolated
EPs in the Brillouin zone (BZ) of a two-dimensional non-Hermitian bilayer
square lattice; protected by symmetry, the isolated EPs only move, merge,
and split in the BZ. The average values of Pauli matrices under the
eigenstate of system Bloch Hamiltonian define a real planar vector field,
the topological defects of which are isolated EPs associated with vortices.
The winding number characterizes the vortices and reveals the topological
properties of the non-Hermitian system. Different topological phases
correspond to different EP configurations, which are unchanged unless
topological phase transition occurs accompanying with the EPs merging or
splitting.
\end{abstract}

\maketitle

\section{Introduction}

Non-Hermitian physics has been extensively investigated in both theoretical
and experimental aspects \cite%
{Moiseyev,Bender,Rotter,Heiss,Cao,Musslimani,AGuo,Peng,LChang,Zhang,Feng17,Longhi17,Huang17,LJinPRL,PTRev,Gupta}%
. In contrast to phase transition in Hermitian systems, exceptional points
(EPs) as non-Hermitian critical points are associated with the coalescence
of eigenstates \cite{Heiss2001,DembowskiPRL2001,DembowskiPRL,DembowskiPRE}.
EPs support many novel phenomena, including power oscillation \cite{Ruter},
square-root frequency dependence \cite{EPSensing1,EPSensing2,EPSensing3},
and gap protected long-range entangled state \cite{LS1}. The eigenstate
flips \cite{Uzdin,BZhen,Doppler,HXu,Hassan,RiemannSheet} and geometric
phases imprinted \cite{Berry,GP,Rotter2008,Huang2013} when encircling EPs
for integer circles reveal the exotic topology of EPs \cite%
{CTChanPRX,CTChanPRX2018,LJin18}. The non-Hermitian topological system
enables novel interface states in quasi-one-dimension and beyond \cite%
{YCHu,Schomerus,Poli,Weimann,TonyLeePRL,CCSymmetry,Leykam,Xiao,Zeuner,Chen,Yuce,LJin,LJinPRA,WangX,Klett,GQLiang,Esaki,Xiao,Lieu,Xiong2018,SChen2018PRA,Borgnia,Zhou,Ezawa,JHu,HZhang,CHe,HChen18,Pan,XWL,Kawabata1,Kawabata2}%
. The topological interface states are classified by two winding numbers
originated from the complex magnitude and the varying direction of an
effective magnetic field \cite{Leykam}. The Chern number and the vorticity
that associated with the eigenstates and the energy band are topological
invariant \cite{LFu}. For the breakdown of conventional bulk-boundary
correspondence, the exotic bulk-edge correspondence is developed \cite%
{WZhong1,WZhong2,ZGong,Kunst}; the importance of chiral inversion symmetry
for the conventional bulk-boundary correspondence is revealed \cite{LJinPRB}%
. Non-Hermitian topological phases and EP rings/surfaces protected by
symmetries are investigated \cite{Budich,Yoshida,Okugawa,ZhouOptica}; the
important application includes the topological lasing \cite%
{Zhao,Parto,Bandres,Harari,St-Jean}.

Recently, topological gapless systems, which are related to diabolic points
(DPs) \cite{MVBerry1984}, have emerged as a new frontier in physics \cite%
{Wan,Yang,Burkov,Xu,Kim,Weng,Huang,Young,Wang,Wang1,Hou,Sama,Liu,Neupane,SXu,Lv,Lu,CTChanPRB,CTChanNP}%
. As a joint of two quantum phases, topological gapless systems have band
structures with band touching points (BTPs) in the momentum space, where
these kinds of nodal points appear as topological defects in an auxiliary
vector field \cite{Xu,Hou}. These points are unremovable due to symmetry
protection until a pair of them meet and annihilate together \cite{JMHouPRB}%
. In general, the EPs of a two-dimensional non-Hermitian system form a loop,
referred as an exceptional ring \cite{Szameit,SHFanEPRing,BZhen,YXu}. A
natural question is whether there exist isolated EPs in the momentum space
which have similar topological characteristics to nodal points in
topological gapless systems.

In this paper, we show that topological nature of BTPs in the Brillouin zone
is determined by the topological defects of a real vector field $\mathbf{F}(%
\mathbf{k})$ mapped from the Bloch Hamiltonian. $\mathbf{F}(\mathbf{k})$ is
composed by the average values of Pauli matrices in non-Hermitian systems in
contrast to an effective magnetic field in Hermitian systems, a winding
number based on which characterizes the vortices at EPs as topological
charges. While, another winding number associated with the energy $\mathbf{%
E(k)}$ characterizes the topology of band coalescence and the chirality of
EPs.\ Moreover, BTPs are protected by system symmetries and their
configurations characterize different topological phases. DPs split to pairs
of EPs when non-Hermiticity is introduced; EPs inherit half-integer vortices
from their parent DPs; topological phase transitions occur associated with
the creation of new configurations when EPs (DPs) merge or split. Topology
of EPs (DPs) is characterized through topological invariant: the winding
numbers $0$ or $\pm 1/2$ ($0$ or $\pm 1$). Our findings are elucidated
through a bilayer square lattice.

\section{Bilayer square lattice}

\begin{figure}[tb]
\includegraphics[ bb=0 80 550 280, width=0.48\textwidth, clip]{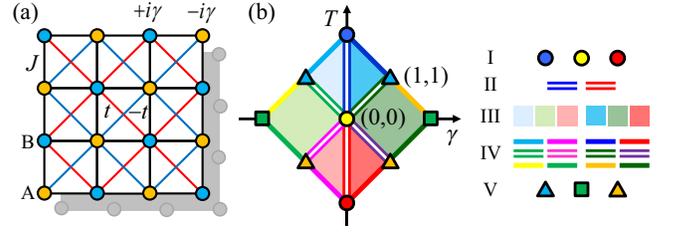}
\caption{(a) Schematic of the bilayer square lattice. The intra-sublattice
hoppings have opposite signs $t$ (red) and $-t$ (blue). The shadowed lattice
indicates the lower layer with $(\protect\gamma ,t)\rightarrow (-\protect%
\gamma ,-t)$ in contrast to the upper layer. (b) Phase diagram of the
bilayer square lattice in the $\protect\gamma $-$T$ plane for $\left\vert
T\pm \protect\gamma \right\vert \leq 2$. Twenty-six topological phases
exhibit five types of winding number distributions of BTPs as listed in
Table \protect\ref{Table I}.}
\label{fig1}
\end{figure}

We consider a tight-binding bilayer square lattice [Fig. \ref{fig1}(a)],
typically describing the dissipative ultracold atomic gas in optical
lattices \cite{DS2008,Cirac2009,Zoller2011,Bardyn2012,Bardyn2013,Budich2015}%
. The Hamiltonian has the form $H=\sum_{\lambda =1}^{2}H_{\lambda }+H_{T}$,
where the intralayer term is
\begin{eqnarray}
H_{\lambda } &=&\sum_{j,l=1}^{N}[J(\left\vert \lambda ,j,l\right\rangle
(\left\langle \lambda ,j+1,l\right\vert +\left\langle \lambda
,j,l+1\right\vert )  \notag \\
&&+t(-1)^{\lambda +j+l}\sum_{\nu =\pm 1}^{{}}\left\vert \lambda
,j,l\right\rangle \left\langle \lambda ,j+1,l+\nu \right\vert +\mathrm{h.c.]}
\notag \\
&&+i\gamma (-1)^{\lambda +j+l}\left\vert \lambda ,j,l\right\rangle
\left\langle \lambda ,j,l\right\vert ,
\end{eqnarray}%
and the interlayer term is $H_{T}=T\left\vert 1,j,l\right\rangle
\left\langle 2,j,l\right\vert +\mathrm{h.c.}$; $\lambda =1$ ($2$) is the
index that labels the upper (lower) layer, and $(j,l)$ is the in-plane site
index. In Fig.\ \ref{fig1}(a), $J$ and $T$ denote the intralayer and
interlayer hoppings; $t$ and $\gamma $ are the diagonal couplings and
staggered losses. Different sublattices have different losses $\gamma _{%
\mathrm{A}}$ and $\gamma _{\mathrm{B}}$ due to the environment interactions.
After a removal of universal loss $(\gamma _{\mathrm{A}}+\gamma _{\mathrm{B}%
})/2$, the losses are equivalently described by the balanced gain and loss $%
\gamma =(\gamma _{\mathrm{A}}-\gamma _{\mathrm{B}})/2$.

The translational symmetry ensures that the Hamiltonian $H$ can be
represented as the summation of a series of $2\times 2$ matrices under the%
\textbf{\ }basis\textbf{\ }$\left\vert \phi (k_{x},k_{y})\right\rangle _{%
\mathrm{A}(\mathrm{B})}=\sum_{j,l=1}^{N}(1/N)e^{i\left( k_{x}j+k_{y}l\right)
}\left\vert \lambda _{\mathrm{A}(\mathrm{B})},j,l\right\rangle $, where $%
k_{x}=2n_{x}\pi /N$, $k_{y}=2n_{y}\pi /N$ with $n_{x}$, $n_{y}\in \left[ 1,N%
\right] $, and states\textbf{\ }$\left\vert \lambda _{\mathrm{A}%
},j,l\right\rangle $ and $\left\vert \lambda _{\mathrm{B}},j,l\right\rangle $%
\textbf{\ }are the position states of sublattices $A$ and $B$ with layer
labels $\lambda _{\mathrm{A}}=[3+\left( -1\right) ^{j+l}]/2$ and $\lambda _{%
\mathrm{B}}=[3-\left( -1\right) ^{j+l}]/2$.

The Hamiltonian $H$ in real space is rewritten as the summation of Bloch
Hamiltonians $h(k_{x},k_{y})$ in the momentum space%
\begin{equation}
H=\sum_{k_{x},k_{y}}h(k_{x},k_{y})\equiv \sum_{k_{x},k_{y}}\mathbf{B}(%
\mathbf{k})\cdot \vec{\sigma}=\sum_{k_{x},k_{y}}B_{x}\sigma _{x}+B_{y}\sigma
_{z},
\end{equation}%
describing an ensemble of non-interacting spin one-half particles in a
two-component complex magnetic field is $\mathbf{B}(\mathbf{k})=\left(
B_{x},B_{y},0\right) $ with%
\begin{equation}
\left\{
\begin{array}{l}
B_{x}=2J(\cos k_{x}+\cos k_{y})+T \\
B_{y}=4t\cos k_{x}\cos k_{y}+i\gamma%
\end{array}%
\right. ,  \label{Bxy}
\end{equation}%
and $\vec{\sigma}=\left( \sigma _{x},\sigma _{z},\sigma _{y}\right) $ is a
vector of Pauli matrices with $\sigma _{x}=\left\vert \phi \right\rangle _{%
\mathrm{A}}\left\langle \phi \right\vert _{\mathrm{B}}+\left\vert \phi
\right\rangle _{\mathrm{B}}\left\langle \phi \right\vert _{\mathrm{A}}$, $%
\sigma _{y}=-i\left\vert \phi \right\rangle _{\mathrm{A}}\left\langle \phi
\right\vert _{\mathrm{B}}+i\left\vert \phi \right\rangle _{\mathrm{B}%
}\left\langle \phi \right\vert _{\mathrm{A}}$, and $\sigma _{z}=\left\vert
\phi \right\rangle _{\mathrm{A}}\left\langle \phi \right\vert _{\mathrm{A}%
}-\left\vert \phi \right\rangle _{\mathrm{B}}\left\langle \phi \right\vert _{%
\mathrm{B}}$ \cite{Note2}. The $\mathbf{B}(\mathbf{k})$ field is invariant
under interchanging of $k_{x}$ and $k_{y}$. The BTPs are at $\left( k_{%
\mathrm{c}x},\pm \pi /2\right) $ and $\left( \pm \pi /2,k_{\mathrm{c}%
y}\right) $, where $\left\vert k_{\mathrm{c}x\left( \mathrm{c}y\right)
}\right\vert =\cos ^{-1}\left[ \left( -T\pm \gamma \right) /\left( 2J\right) %
\right] $; the BTPs are EPs (DPs) for $\gamma \neq 0$ ($\gamma =0$).

\section{Symmetry}

Discrete symmetries play a crucial role in characterizing topological phases
\cite{Ludwig}. Here, $t$ breaks the inversion symmetry and $\gamma $ breaks
the time-reversal symmetry of the bilayer square lattice, but $H$ has a
chiral symmetry $\chi H\left( J,T,\gamma ,t\right) \chi ^{-1}=-H\left(
J,T,\gamma ,t\right) $. The chiral operator $\chi =UC_{4}$ is a combination
of a $Z_{2}$ gauge transformation $U$ and a $90$ degrees rotation $C_{4}$.
The $Z_{2}$ gauge transformation $U\left\vert \lambda _{\mathrm{A}(\mathrm{B}%
)},j,l\right\rangle =\left( -1\right) ^{\lambda +j+l}\left\vert \lambda _{%
\mathrm{A}(\mathrm{B})},j,l\right\rangle $ changes the system Hamiltonian $%
H\left( J,T,\gamma ,t\right) $ into $H\left( -J,-T,\gamma ,t\right) $. The
chiral symmetry instead of time-reversal symmetry or inversion symmetry
ensures the Bloch Hamiltonian including only two Pauli matrices \cite{Sama}.

The mirror-reflection in the $x$ and $y$ directions ($M_{x}$ and $M_{y}$),
the translation of one site in the $x$ and $y$ directions ($T_{x}$ and $%
T_{y} $), the $90$ degrees rotation ($C_{4}$), and the upper and lower layer
interchange ($\Lambda $) all change the Hamiltonian $H\left( J,T,\gamma
,t\right) $ into $H\left( J,T,-\gamma ,-t\right) $. Therefore, $H$ is
invariant under a combination of any two of the above listed operations and
leads to (Appendix A)%
\begin{equation}
h(k_{x},k_{y})=h(\pm k_{x(y)},\pm k_{y(x)}).
\end{equation}%
Therefore, the number of EPs (DPs) in the bilayer lattice is protected by
these symmetries, being invariant unless topological phase transition occurs
when EPs (DPs) merge or split in the non-Hermitian (Hermitian) bilayer
lattice.

Notably, the non-Hermitian bilayer square lattice $H\left( J,T,\gamma
,t\right) $ does not have either $\mathcal{PT}$ symmetry or $\mathcal{CT}$
symmetry (charge-conjugation symmetry in Ref.~\cite{CCSymmetry}) in the
situation that $t\neq 0$. The operators $\mathcal{C}$ is defined as $%
\mathcal{C}\left\vert \lambda _{\mathrm{A}(\mathrm{B})},j,l\right\rangle
=\left( -1\right) ^{0(1)}\left\vert \lambda _{\mathrm{A}(\mathrm{B}%
)},j,l\right\rangle $ \cite{LS2}, $\mathcal{T}$ is defined as $\mathcal{T}i%
\mathcal{T}^{-1}=-i$, and $\mathcal{P}$ is defined as $\mathcal{P}\left\vert
\lambda _{\mathrm{A}(\mathrm{B})},j,l\right\rangle =\left\vert \lambda _{%
\mathrm{A}(\mathrm{B})},N+1-j,l\right\rangle $ or $\left\vert \lambda _{%
\mathrm{A}(\mathrm{B})},j,N+1-l\right\rangle $. In the absence of diagonal
coupling ($t=0$), the system satisfies both $\mathcal{PT}$ symmetry [$\left(
\mathcal{PT}\right) H\left( J,T,\gamma ,t\right) \left( \mathcal{PT}\right)
^{-1}=H\left( J,T,\gamma ,t\right) $] and $\mathcal{CT}$ symmetry [$\left(
\mathcal{CT}\right) H\left( J,T,\gamma ,t\right) \left( \mathcal{CT}\right)
^{-1}=-H\left( J,T,\gamma ,t\right) $] \cite{LS2}. $\mathcal{PT}$ symmetry
leads to the spectrum being real or being conjugate pairs. $\mathcal{CT}$
symmetry requires the eigen energies appearing in pairs with opposite real
parts and identical imaginary parts. Thus, the spectrum under $\mathcal{PT}$
symmetry and $\mathcal{CT}$ symmetry is purely real or purely imaginary.

The system is inversion symmetric and time-reversal symmetric when $t=\gamma
=0$ \cite{LFu,LFuPRB}. The inversion symmetry changes the sign of the
momentum and interchanges the upper and lower layers.

\begin{table}[tbp]
\caption{Five types of BTP distributions. }
\label{Table I}\renewcommand\arraystretch{1} 
\par
\begin{tabular}{c|ccccc}
\hline\hline
$w_{\mathrm{I,II}}$ & \textrm{I} & \textrm{II} & \textrm{III} & \textrm{IV}
& \textrm{V} \\
$\left\{ 0,\pm 1/2,\pm 1\right\} $ & $\left\{ 4,0,0\right\} $ & $\left\{
0,0,8\right\} $ & $\left\{ 0,16,0\right\} $ & $\left\{ 4,8,0\right\} $ & $%
\left\{ 8,0,0\right\} $ \\ \hline\hline
\end{tabular}%
\end{table}

\section{Topological invariants of EP}

$h(k_{x},k_{y})$ includes only two Pauli matrices and commutes with the
third one, we employ a planar vector field $\mathbf{F}(\mathbf{k})=\left(
F_{x},F_{y}\right) $ to characterize the topology of $h(k_{x},k_{y})$. The
zero energy band degeneracy $E_{\pm }=\pm \sqrt{B_{x}^{2}+B_{y}^{2}}=0$
implies that the expectation value of $h(k_{x},k_{y})$ under its eigenstate
is zero, i.e., $\left\langle h(k_{x},k_{y})\right\rangle =B_{x}\left\langle
\sigma _{x}\right\rangle +B_{y}\left\langle \sigma _{z}\right\rangle =0$;
therefore, $B_{x}=B_{y}=0$ ($\left\langle \sigma _{x}\right\rangle
=\left\langle \sigma _{z}\right\rangle =0$) at DPs (EPs) in Hermitian
(non-Hermitian) system. The expectation values of the two Pauli matrices
compose the planar vector field%
\begin{equation}
\mathbf{F}(\mathbf{k})=\left( \left\langle \sigma _{x}\right\rangle
,\left\langle \sigma _{z}\right\rangle \right) ,
\end{equation}%
for non-Hermitian system instead of $\mathbf{F}(\mathbf{k})=\left(
B_{x},B_{y}\right) $ for Hermitian system \cite{Sama,Hou}. EPs are protected
by symmetry, exhibiting similar behaviors as DPs in Hermitian lattices and
being topologically stable. Moreover, EPs can map to the topological defects
of the manifold $\mathbf{F}(\mathbf{k})$, at which $\mathbf{F}(\mathbf{k})$
approaches zero with vortex or antivortex structures.

A topological invariant characterizes the topological property of $\mathbf{F}%
(\mathbf{k})$
\begin{equation}
w_{\mathrm{I}}=\oint_{\mathrm{C}}\frac{\mathrm{d}\mathbf{k}}{2\pi }\left(
\hat{F}_{x}\nabla \hat{F}_{y}-\hat{F}_{y}\nabla \hat{F}_{x}\right) ,
\label{Winding N}
\end{equation}%
where $\hat{F}_{x(y)}=F_{x(y)}/\sqrt{F_{x}^{2}+F_{y}^{2}}$ and $\nabla
=\partial /\partial \mathbf{k}$. The integral is performed in the $k_{x}$-$%
k_{y}$ plane along a closed loop $\mathrm{C}$; $2\pi w_{\mathrm{I}}$
accounts the varying direction of vector field $\mathbf{F}(\mathbf{k})$. The
(both eigenstates yielding identical) winding number $w_{\mathrm{I}}$
describes the vortices with topological charges \cite{LFu,YXu} at EPs and is
in parallel to the winding number that originated from the generalized Berry
phase \cite{YXu,CTChanPRB,Leykam}.

Moreover, the topology of BTP can be determined by winding numbers $w_{%
\mathrm{I}}$\ via the planar vector fields $\mathbf{F(k)}$. By means of $w_{%
\mathrm{I}}$, the BTP is classified into four types:

(i) Semi-Dirac point: $H(\mathbf{k})$\ is Hermitian and semi-Dirac point
possesses an anisotropic dispersion of linear and quadratic dispersions
along two orthogonal directions \cite{Dietl,Banerjee,JKim,LDu,QChen}. It has
winding number $w_{\mathrm{I}}=0$\ and can split into two Dirac points ($w_{%
\mathrm{I}}=\pm 1$) when changing parameters without breaking any symmetries
\cite{KSunPRL2009}.

(ii) Dirac point: $H(\mathbf{k})$\ is also Hermitian and the dispersion is
linear along all directions around the Dirac point, but the vector field $%
\mathbf{F(k)}$\ owns a vortex ($w_{\mathrm{I}}=+1$) or an antivortex ($w_{%
\mathrm{I}}=-1$) with opposite topological charges \cite%
{Montambaux,MontambauxPRL,Esslinger}. In the presence of non-Hermiticity,
each Dirac point splits into two normal EPs with identical half-integer
vortices as in (iii) \cite{Keck,Seyranian,ZhenScience}; in contrast, the
Dirac point can split into two Weyl points in Hermitian systems with
breaking inversion or time-reversal symmetry \cite%
{LFuNC2012,LinNC2016,LFuPRB,Young2012,Yang2014,BalentsPRL,Balents,LLuScience,BQLv}%
.

(iii) Normal EP: The winding numbers of normal EPs are $w_{\mathrm{I}}=\pm
1/2$; it indicates that EPs inherit one-half vortices from their parent
Dirac points with $w_{\mathrm{I}}=\pm 1$ \cite{Hou,Sama}. The dispersion is
square-root along any direction near the normal EPs.

(iv) Hybrid EP: two normal EPs with opposite vortices can merge into a
hybrid EP. In this situation, the winding number of EP is changed from $\pm
1/2$\ to $0$.\ Different from that of a semi-Dirac point \cite%
{Dietl,Banerjee,JKim,LDu,QChen}, hybrid EP also has an anisotropic
dispersion with square-root (linear) dispersions along two orthogonal
directions \cite{LFu}.

While, in parallel to $w_{\mathrm{I}}$, another winding number is defined as%
\begin{equation}
w_{\mathrm{II}}=\oint_{\mathrm{C}}\frac{\mathrm{d}\mathbf{k}}{2\pi }\left(
\hat{E}_{x}\nabla \hat{E}_{y}-\hat{E}_{y}\nabla \hat{E}_{x}\right) ,
\end{equation}%
with $\mathbf{E}(\mathbf{k})=\left( E_{x},E_{y}\right) $ being the band
energy $E=\pm \sqrt{B_{x}^{2}+B_{y}^{2}}$ in the complex plane, and $\hat{E}%
_{x}=$\textrm{Re(}$E$\textrm{)/}$|E|$, $\hat{E}_{y}=$\textrm{Im(}$E$\textrm{%
)/}$|E|$, which characterizes the topology of EPs on another aspect. Both
two bands yield identical $w_{\mathrm{II}}$ due to chiral symmetry. $w_{%
\mathrm{II}}$ differs from $w_{\mathrm{I}}$ and characterizes the topology
of band coalescence \cite{Leykam,LFu}, reveals the vorticity of complex
Riemann surface band structure \cite{Doppler,RiemannSheet,HXu}, and reflects
the chirality of normal EPs \cite{GP,Rotter2008,Huang2013,CTChanPRX,Note}.
Each of the two bands accumulates Berry phase $\pm \pi $ when encircling EPs
for two circles; the $\pm $ sign in front determines the opposite
chiralities of EPs \cite{DembowskiPRL,DembowskiPRE,GP,Uzdin,Huang2013} as a
result of the square-root type Riemann sheet. The merge of EPs with opposite
chiralities creates DP or hybrid EP \cite{CTChanPRX,LFu}. The winding number
is $w_{\mathrm{II}}=\pm 1/2$ ($0$) for normal EP (DP and hybrid EP).

\section{Topological phases}

\begin{figure}[tb]
\includegraphics[ bb=40 10 581 541, width=0.49\textwidth, clip]{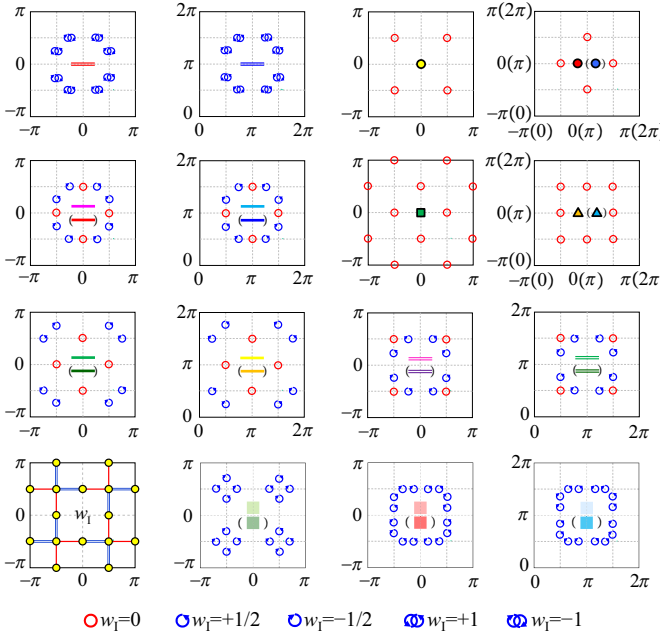}
\caption{Schematic of BTP configurations of $w_{\mathrm{I}}$ for the
topological phases in Fig. \protect\ref{fig1}(b). The horizontal (vertical)
axis is $k_{x}$ ($k_{y}$). The blue arrows and red circles represent $w_{%
\mathrm{I}}=\pm 1/2$, $\pm 1$ and $0$. $w_{\mathrm{I}}$ of the Hermitian
(non-Hermitian) bilayer lattice is shown in the bottom left corner; the red
single and blue double lines represent $w_{\mathrm{I}}=+1$ and $-1$ ($w_{%
\mathrm{I}}=+1/2$ and $-1/2$), and the yellow filled circles represent $w_{%
\mathrm{I}}=0$ ($w_{\mathrm{I}}=0$). }
\label{fig2}
\end{figure}

\begin{figure}[tb]
\includegraphics[ bb=25 10 588 541, width=0.510\textwidth, clip]{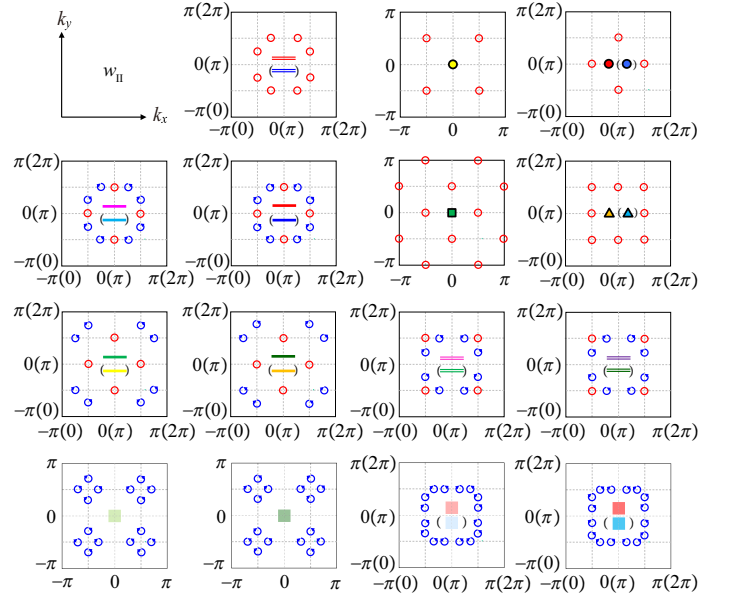}
\caption{Schematic of BTP configurations of $w_{\mathrm{II}}$ for the
topological phases in Fig. \protect\ref{fig1}(b), the corresponding BTP
configurations of $w_{\mathrm{I}}$ is in Fig.~\protect\ref{fig2}.}
\label{fig3}
\end{figure}

$H$ is invariant under arbitrary two combinations of the mirror reflection,
the translation, the $90$ degrees rotation, and the layer interchange.
Symmetries protect the Bloch Hamiltonian $h(k_{x},k_{y})$, being invariant
under substitution $\left( k_{x},k_{y}\right) \rightarrow (\pm k_{x(y)},\pm
k_{y(x)})$, and also protect the number of BTPs (DPs and EPs) with vortex
structures. The BTPs can move but cannot be removed as system parameters
varying except when they merge or split associated with a topological phase
transition and the creation of a new BTP configuration. The BTP
configuration (Figs. \ref{fig2} and \ref{fig3}) characterizes the
topological phases of bilayer square lattice, while the phase diagram is
depicted in Fig. \ref{fig1}(b).

The system is Hermitian at $\gamma =0$, being time-reversal symmetric
without inversion symmetry ($t\neq 0$). The time-reversal symmetry requires
DPs at $\pm k$ with identical winding number \cite{Young,LinNC2016}. Eight
DPs may exist due to the symmetry protection. They locate and move along the
$\left\vert k_{x(y)}\right\vert =\pi /2$\ lines; their locations satisfy a
fourfold rotational symmetry ($C_{4}$ symmetry) with respect to the axis
that perpendicular to the $k_{x}$-$k_{y}$ plane, being
mirror-reflection-symmetric about $k_{x(y)}=0$ and $k_{x}=\pm k_{y}$ \cite%
{Balents}. Eight DPs merge into four at $\left\vert T\right\vert =0$ and $2J$
\cite{Young,LFuNC2012}. Top panel of Fig. \ref{fig2} shows the DP
configurations.

At $\gamma \neq 0$, each DP splits into two normal EPs and the corresponding
Bloch Hamiltonian $h(k_{x},k_{y})$ is defective \cite%
{Keck,Seyranian,ZhenScience}, generating sixteen EPs at maximum. $%
h(k_{x},k_{y})$ has chiral symmetry, which requires the EPs ($E_{\pm }=0$)
on the lines of $k_{x(y)}=\pm \pi /2$ if $t\neq 0$. Although EPs are
independent of $t$, nonzero $t$ is crucial for the existence of isolated
EPs. For special system parameters, the hybrid EPs appear at high symmetric
points in the Brillouin zone, where normal EPs with the opposite charges
merge and the number of EPs reduces.

$w_{\mathrm{I}}$ characterizes the topological features of BTPs as the
topological defects of planar vector field $\mathbf{F}(\mathbf{k})$. The
value of $w_{\mathrm{I}}$ is given in Fig. \ref{fig2}. Distinct topological
phases are created when BTPs merge or split at topological phase transition.
Topological phases being reflection symmetric about $\gamma =0$ has
identical $w_{\mathrm{I}}$ configurations, exhibiting seventeen
configurations of $w_{\mathrm{I}}$ [Fig. \ref{fig2}].

In Fig. \ref{fig2}, BTPs with $w_{\mathrm{I}}=0$ are semi-Dirac points on
the top panel and are hybrid EPs otherwise. The dispersion is linear along
the $k_{x(y)}$ ($k_{x}=\pm k_{y}$) direction near the semi-Dirac points on
the $k_{x(y)}=\pm \pi /2$ ($k_{x}=\pm k_{y}$) line. The dispersion along the
$\lambda _{1}k_{x}=\lambda _{2}k_{y}$ ($\lambda _{1}k_{x}=-\lambda _{2}k_{y}$%
) direction is square-root (linear) near the hybrid EPs $\left( \lambda
_{1}\pi /2,\lambda _{2}\pi /2\right) $, where $\lambda _{1,2}=\pm 1$. The
dispersion along the $k_{x}=0$ ($k_{y}=0$) direction is square-root (linear)
near the hybrid points $\left( 0,\pm \pi /2\right) $ and $\left( \pi ,\pm
\pi /2\right) $.

\begin{figure}[tbp]
\includegraphics[ bb=0 146 345 620, width=0.49\textwidth, clip]{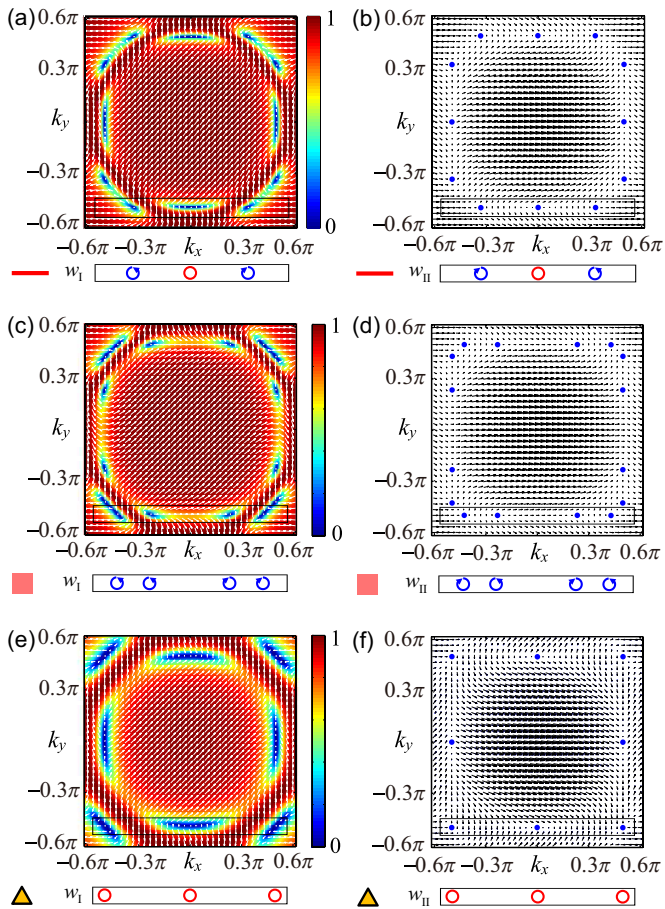}
\caption{(a) The planar vector field in the momentum space $\mathbf{F}(%
\mathbf{k})=\left( F_{x},F_{y}\right) =\left( \left\langle \protect\sigma %
_{x}\right\rangle ,\left\langle \protect\sigma _{z}\right\rangle \right) $
is shown by the white arrows on the top, and the background density plot is $%
F_{x}^{2}+F_{y}^{2}$. (b) $\mathbf{E}{(\mathbf{k})}=\left(
E_{x},E_{y}\right) =\left( \mathrm{Re}(E),\mathrm{Im}(E)\right) $. The plots
are for $E_{+}$ at system parameters $\protect\gamma =J/2,T=-3J/2$. The
vortices in the rectangles are indicated by the winding numbers $w_{\mathrm{%
I(II)}}=+(-)\{+1/2,0,-1/2\}$ from the left to the right. (c, d) The system
parameters are $\protect\gamma =J/2,T=-J$, and (e, f) $\protect\gamma =T=-J$%
. The winding numbers of EPs inside the rectangles are schematically
illustrated at the bottom of each plot.}
\label{fig4}
\end{figure}

Three typical types of vortices for the planar vector fields $\mathbf{F}%
\left( \mathbf{k}\right) $ and $\mathbf{E}\left( \mathbf{k}\right) $ in the
bilayer lattice are shown in Fig. \ref{fig4}. The EP configurations are
depicted, including both the zero and half-integer vortices of $w_{\mathrm{%
I,II}}=0$ and $\pm 1/2$ for the EPs. Figures \ref{fig4}(a) and \ref{fig4}(b)
include eight one-half vortices (normal EP $w_{\mathrm{I,II}}=\pm 1/2$) and
four zero vortices (hybrid EP $w_{\mathrm{I,II}}=0$). The three EPs along $%
k_{y(x)}=-(+)\pi /2$ from $k_{x(y)}=-\pi $ to $k_{x(y)}=+\pi $ have $w_{%
\mathrm{I}}=\{1/2,0,-1/2\}$; in contrast, $w_{\mathrm{II}}=\{-1/2,0,+1/2\}$.
Figures \ref{fig4}(c) and \ref{fig4}(d) include sixteen one-half vortices
(normal EP $w_{\mathrm{I,II}}=\pm 1/2$). Figures \ref{fig4}(e) and \ref{fig4}%
(f) include eight zero vortices (hybrid EP $w_{\mathrm{I,II}}=0$). In
essential, the winding numbers $w_{\mathrm{I}}$ and $w_{\mathrm{II}}$
reflect the varying directions of planar vector fields $\mathbf{F}\left(
\mathbf{k}\right) $ and $\mathbf{E}\left( \mathbf{k}\right) $, respectively;
they both capture two topological aspects of EPs and provide an overall
perspective of the topological properties of non-Hermitian system. $w_{%
\mathrm{I}}$\ reflects the topological properties of system: EPs naturally
inherit fractional topological charges from DPs in the Hermitian situation; $%
w_{\mathrm{I}}$ is a topological invariant that equivalent to the winding
number originated from the non-Hermitian generalization of Berry phase that
characterizing the varying direction of the effective magnetic field \cite%
{Leykam,LFu,YXu}.

$w_{\mathrm{II}}$ characterizes the chirality of EPs. When $\gamma
\rightarrow -\gamma $, the band energy vector becomes\ $\mathbf{E}^{^{\prime
}}\mathbf{(k)}=\left( E_{x},-E_{y}\right) $\ with $w_{\mathrm{II}}\left(
-\gamma \right) =-w_{\mathrm{II}}\left( \gamma \right) $. But $w_{\mathrm{I}%
} $ is unchanged with $w_{\mathrm{I}}\left( -\gamma \right) =w_{\mathrm{I}%
}\left( \gamma \right) $. The phases symmetric about the $T$-axis in phase
diagram [Fig. \ref{fig1}(b)] are two different phases characterized by
opposite $w_{\mathrm{II}}$. Thus, all twenty-six BTP configurations of $w_{%
\mathrm{II}}$\ distinguish the different topological phases. Compared with
the BTP configurations of $w_{\mathrm{I}}$, the additional nine
configurations are attributed to the chirality of EPs that relevant to the
positive or negative value of $\gamma $. The BTP configurations of $w_{%
\mathrm{II}}$ are shown in Fig. \ref{fig3} as a supplementary and a
comparison to the BTP configurations of $w_{\mathrm{I}}$ of Fig. \ref{fig2}.
All different phases are distinguishable in the $w_{\mathrm{II}}$ BTP
configurations, where the chirality information of EPs are included \cite%
{DembowskiPRL2001}.

The diagonal coupling term $t$ breaks the inversion symmetry and plays a key
role in the existence of isolated topologically stable EPs. When $t=0$, the
EPs are no longer isolated points but rings at $\cos k_{\mathrm{c}x}+\cos k_{%
\mathrm{c}y}=\left( -T\pm \gamma \right) /\left( 2J\right) $ \cite%
{XuPRA,DWZPRA,Szameit,SHFanEPRing,BZhen,LS2,Budich,Yoshida,Okugawa}.
However, when $t=0$\ and $\left\vert \gamma \right\vert =\left\vert
T-4J\right\vert $ ($\left\vert T+4J\right\vert $), the EP rings \cite{BZhen}
reduce to isolated points at $\left( k_{\mathrm{c}x},k_{\mathrm{c}y}\right) $
of $k_{\mathrm{c}x}=k_{\mathrm{c}y}=\pm \pi $ ($0$) with trivial topological
features $w_{\mathrm{I}}=w_{\mathrm{II}}=0$; the real or imaginary part of
spectrum in the momentum space possesses a Dirac-cone band structure for $%
T>4J$ or $T<4J$ ($T<-4J$ or $T>-4J$) at $J>0$; EP vanishes at $\left\vert
\gamma \right\vert >\max \{\left\vert T+4J\right\vert ,\left\vert
T-4J\right\vert \}\ $or $\left\vert \gamma \right\vert <\min \{\left\vert
T+4J\right\vert ,\left\vert T-4J\right\vert \}$ if $\left\vert T\right\vert
>4J$. Recently, the topological features of EP rings are revealed \cite%
{YXu,CerjanPRB}.

\section{Discussion and Conclusion}

Ultracold atomic gas in optical lattices \cite%
{DS2008,Cirac2009,Zoller2011,Bardyn2012,Bardyn2013,Budich2015,SLZhu,Bloch,Zoller,XJLiuPRL,BlochPRL,LiuPRL2014,Goldman,JWPan}%
, photonic crystals \cite{PC,Hafezi2016}, and coupled resonators \cite%
{Leykam,Hafezi2011,KFang,Leykam2,LLuRev,SHFanNC2016,Zhao,Parto,ZhenScience}
provide versatile optical platforms for the study of topological physics; the dissipation and radiation in optical systems are
non-Hermitian and thus can be considered as the possible candidate for the experimental realization to the
non-Hermitian bilayer square lattice.
The fine tuned system parameters are particularly beneficial for the
realization of topological systems. Several pioneering experimental and
theoretical possibilities have been devoted to ultracold atoms in hexagonal
and square optical lattices \cite%
{Baranov,Soltan-Panahi,Tobias,Gross,Barredo,Yoo}. A particular feature of
cold atoms in optical lattice is the tunability of the tunneling strengths. One example is the study of topological quantum phase transition in a
bilayer square lattice describing ultracold atomic gases in optical
lattices. Trapping cold $^{6}$\textrm{Li} or $^{40}$\textrm{K }atoms in a
dissipative spin-dependent optical lattice is a candidate for realizing the
bilayer square lattice \cite{YXu,Sama,DS2008,Cirac2009,Bardyn2013}, and
next-nearest-neighbour (diagonal) coupling is introduced by applying
additional laser beams. Moreover, the other possible
realizations are the coupled resonator waveguide lattice \cite%
{Hafezi2011,KFang,LLuRev,Leykam,Leykam2,SHFanNC2016,Zhao,Parto,ZhenScience}
and the photonic crystals \cite{PC,Hafezi2016}. The $\pi $ phase difference
between the diagonal couplings $t$ of sublattices $A$ and $B$ is controlled
by the optical path length of auxiliary linking resonators (waveguides).

In summary, the symmetry protected isolated EPs in a non-Hermitian system
are studied, which differs from the symmetry protected EP rings and EP
surfaces \cite{Budich,Yoshida,Okugawa}. A real valued vector field $\mathbf{F%
}(\mathbf{k})$ is defined by the average values of Pauli matrix. The
topological defects of the field are the BTPs associated with the vortices
characterized by winding number $w_{\mathrm{I}}$, which reflects the
topological properties of the system. The EPs own fractional topological
charges that inherit from their parent DPs in Hermitian systems. The type
and number of BTPs is protected by discrete symmetries and only changes at
the annihilation or split of vortex and antivortex accompanied by a
topological phase transition. A pair of half-integer topological charged EPs
can merge into a hybrid EP with vanishing topological charge or a DP with
integer topological charge zero or one. Another winding number $w_{\mathrm{II%
}}$ characterizes the chirality of EPs and the Riemann sheet band structure.
$w_{\mathrm{I}}$\ and $w_{\mathrm{II}}$\ of EPs provide an overall
perspective of the topological properties of the non-Hermitian system and
the BTP configurations outline distinct topological phases. Our findings
deepen the understanding of symmetry protected non-Hermitian topological
phases and may stimulate interest in chasing for topologically stable EPs in
other physical systems.

\acknowledgments We acknowledge the support of National Natural Science
Foundation of China (Grants No. 11374163 and No. 11605094).

\appendix

\section*{Appendix A: Symmetry}

\renewcommand{\theequation}{A\arabic{equation}} \setcounter{equation}{0} The
mirror-reflection in the $x$ and $y$ directions ($M_{x}$ and $M_{y}$), the
translation of one site in the $x$ and $y$ directions ($T_{x}$ and $T_{y} $%
), the $90$ degrees rotation ($C_{4}$), and the upper and lower layer
interchange ($\Lambda $) all change the non-Hermitian bilayer square lattice
Hamiltonian $H\left( J,T,\gamma ,t\right) $ into $H\left( J,T,-\gamma
,-t\right) $. Therefore, $H$ is invariant under a combination of any two of
the above listed operations. For example, a combination operator $\Theta
_{1}=M_{x}M_{y}$ leads to $\Theta _{1}H\Theta _{1}^{-1}=H$; under this
symmetry operation, the Bloch Hamiltonian in the momentum space changes into
$\Theta _{1}h(k_{x},k_{y})\Theta _{1}^{-1}=h(-k_{x},-k_{y})$. Other symmetry
$\Theta _{2}=\Lambda M_{x}$ leads to $\Theta _{2}H\Theta _{2}^{-1}=H$; thus,
we have $\Theta _{2}h(k_{x},k_{y})\Theta _{2}^{-1}=h(-k_{x},k_{y})$.
Similarly, set $\Theta _{3}=\Lambda M_{y}$, we have $\Theta
_{3}h(k_{x},k_{y})\Theta _{3}^{-1}=h(k_{x},-k_{y})$; set $\Theta
_{4}=M_{x}C_{4}$, we have $\Theta _{4}h(k_{x},k_{y})\Theta
_{4}^{-1}=h(k_{y},k_{x})$; set $\Theta _{5}=M_{y}C_{4}$, we have $\Theta
_{5}h(k_{x},k_{y})\Theta _{5}^{-1}=h(-k_{y},-k_{x})$; set $\Theta
_{6}=\Lambda C_{4}M_{x}M_{y}$, we have $\Theta _{6}h(k_{x},k_{y})\Theta
_{6}^{-1}=h(k_{y},-k_{x})$; set $\Theta _{7}=C_{2}$, we have $\Theta
_{7}h(k_{x},k_{y})\Theta _{7}^{-1}=h(-k_{x},-k_{y})$. These discrete
symmetries of the non-Hermitian bilayer square lattice lead to%
\begin{equation}
h(k_{x},k_{y})=h(\pm k_{x(y)},\pm k_{y(x)}).
\end{equation}%
Therefore, the number of EPs (DPs) in the bilayer lattice is protected by
these symmetries, being invariant unless the topological phase transition
occurs when EPs (DPs) merge in the non-Hermitian (Hermitian) bilayer lattice.

\begin{figure*}[tbh]
\includegraphics[ bb=10 5 525 160, width=1.0\textwidth, clip]{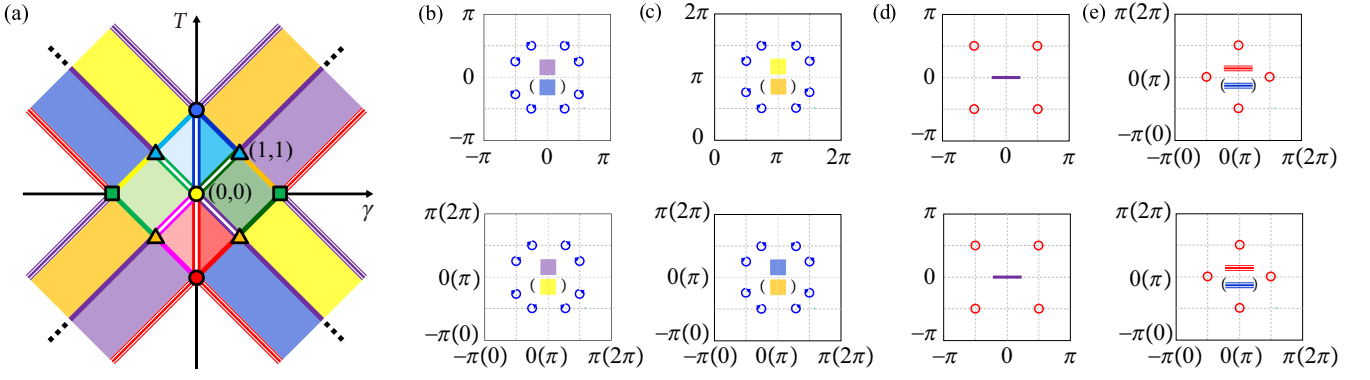}
\caption{ (a) The phase diagram in the parameter space $\protect\gamma $-$T$%
. $16$ ($8$) EPs appear in the colored area inside (outside) the region $%
\left\vert T\pm \protect\gamma \right\vert \leq 2$, and the system is gapped
in the white region. Excepted for the whole twenty-six BTP configurations
shown in Figs. \protect\ref{fig2} and \protect\ref{fig3}, other BTP
configurations outside the region $\left\vert T\pm \protect\gamma %
\right\vert \leq 2$ are illustrated in (b-e). The upper (lower) panel is the
configurations of $w_{\mathrm{I}}$ ($w_{\mathrm{II}}$).}
\label{figEPs}
\end{figure*}

\section*{Appendix B: BTP configuration}

\renewcommand{\theequation}{B\arabic{equation}} \setcounter{equation}{0} The
complete phase diagram of the bilayer square lattice is illustrated in Fig. %
\ref{figEPs}(a); in contrast to Fig. 1(b), Fig. \ref{figEPs}(a)
distinguishes all different types of BTP configurations in the system
parameter space $\gamma $-$T$. In Figs. \ref{figEPs}(b)-\ref{figEPs}(e), the
system parameters are in the colored areas outside the region $\left\vert
T\pm \gamma \right\vert \leq 2$ as a complement of Figs. \ref{fig2} and \ref%
{fig3}; notably, all the EPs are still located at $\left( k_{\mathrm{c}%
x},\pm \pi /2\right) $ and $\left( \pm \pi /2,k_{\mathrm{c}y}\right) $ with $%
\left\vert k_{\mathrm{c}x\left( \mathrm{c}y\right) }\right\vert =\cos ^{-1}%
\left[ \left( -T\pm \gamma \right) /\left( 2J\right) \right] $. These
topological phases are not discussed in the main text due to the simplicity
of this situation. In this situation, eight normal EPs with $w_{\mathrm{I}%
}=\pm 1/2$ exist as shown on the upper panel of Figs. \ref{figEPs}(b) and %
\ref{figEPs}(c), and their mergence creates the new configurations as shown
on the upper panel of Figs. \ref{figEPs}(d) and \ref{figEPs}(e), where four
hybrid EPs appear. The BTP configurations for $w_{\mathrm{II}}$ outside the
region $\left\vert T\pm \gamma \right\vert \leq 2$ are depicted on the lower
panel of Figs. \ref{figEPs}(b)-\ref{figEPs}(e).

\section*{Appendix C: Dispersion near the BTPs}

\renewcommand{\theequation}{C\arabic{equation}} \setcounter{equation}{0} In
this Appendix, we discuss the dispersions near the BTPs (EPs and DPs) with
different winding numbers. Here we only discuss the dispersion around EPs at
$\left( k_{\mathrm{c}x},\pm \pi /2\right) $, and the situations of $\left(
\pm \pi /2,k_{\mathrm{c}y}\right) $ can be obtained by interchanging $k_{x}$
and $k_{y}$. Set $q_{x}=k_{x}-k_{\mathrm{c}x}$ and $q_{y}=k_{y}-\pi /2 $,
the Taylor expansion of the $\mathbf{B}(\mathbf{k})$ field to the second
order near $\left( k_{\mathrm{c}x},\pm \pi /2\right) $ yields%
\begin{equation}
\left\{
\begin{array}{l}
B_{x}\approx 2J\left[ \left( 1-q_{x}^{2}/2\right) \cos k_{cx}-q_{x}\sin
k_{cx}\mp q_{y}\right] +T \\
B_{y}\approx \mp 4tq_{y}\left[ \left( 1-q_{x}^{2}/2\right) \cos
k_{cx}-q_{x}\sin k_{cx}\right] +i\gamma%
\end{array}%
\right. ,  \label{BxBy}
\end{equation}%
and the dispersion is%
\begin{equation}
E_{\pm }=\pm \sqrt{B_{x}^{2}+B_{y}^{2}}.
\end{equation}%
In the following, we discuss the dispersion near the EPs and DPs in detail.

Firstly, we consider the dispersion near normal EPs with $w_{\mathrm{I}}=\pm
1/2$. Near $\left( k_{\mathrm{c}x},\pm \pi /2\right) $ with $k_{\mathrm{c}%
x}\neq 0$, $\pi $, $\pm \pi /2$, we approximately have the dispersion along
the $k_{y}=\xi k_{x}$ ($k_{x}=-\xi k_{y}$) direction as $E_{+}\left(
k_{y}=\xi k_{x}\right) \approx 2\sqrt{-J\left( T+2J\cos k_{\mathrm{c}%
x}\right) \left( \sin k_{\mathrm{c}x}\pm \xi \right) \mp 2i\gamma t\xi \cos
k_{\mathrm{c}x}}\sqrt{q_{x}}$ and $E_{+}\left( k_{x}=-\xi k_{y}\right)
\approx 2\sqrt{J\left( T+2J\cos k_{\mathrm{c}x}\right) \left( \xi \sin k_{%
\mathrm{c}x}\mp 1\right) \mp 2i\gamma t\cos k_{\mathrm{c}x}}\sqrt{q_{y}}$,
where $\xi $ is real. Notably, the dispersion along any direction is
square-root near normal EPs with $w_{\mathrm{I}}=\pm 1/2$.

Secondly, we consider the dispersion near hybrid EPs with $w_{\mathrm{I}}=0$%
. The hybrid points with $w_{\mathrm{I}}=0$ at $k_{\mathrm{c}x}=0,\pi ,\pm
\pi /2$ have anisotropic dispersion, being square-root in one direction and
linear in the perpendicular direction \cite{LFu}. For $k_{\mathrm{c}x}=\pm
\pi /2$ (i.e., $T\pm \gamma =0$), we approximately have $\mathbf{B}(\mathbf{k%
})$ field near $\left( \lambda _{1}\pi /2,\lambda _{2}\pi /2\right) $ as%
\begin{equation}
\left\{
\begin{array}{l}
B_{x}\approx -2J\left( \lambda _{1}q_{x}+\lambda _{2}q_{y}\right) +T \\
B_{y}\approx 4\lambda _{1}\lambda _{2}tq_{x}q_{y}+i\gamma%
\end{array}%
\right. ,
\end{equation}%
where $\lambda _{1},\lambda _{2}=\pm $. The dispersion along the $\lambda
_{1}k_{x}=\lambda _{2}k_{y}$ ($\lambda _{1}k_{x}=-\lambda _{2}k_{y}$)
direction is square-root (linear) as%
\begin{eqnarray}
&&E_{+}\left( \lambda _{1}k_{x}=\lambda _{2}k_{y}\right) \approx 2\sqrt{%
-2\lambda _{1}JT}\sqrt{q_{x}}, \\
&&E_{+}(\lambda _{1}k_{x}=-\lambda _{2}k_{y})\approx 2\sqrt{-2i\gamma t}%
\left\vert q_{x}\right\vert .
\end{eqnarray}%
For $k_{\mathrm{c}x}=0$ (i.e., $T\pm \gamma =-2J$), we approximately have $%
\mathbf{B}(\mathbf{k})$ field near $\left( 0,\pm \pi /2\right) $ as%
\begin{equation}
\left\{
\begin{array}{l}
B_{x}\approx -J\left( q_{x}^{2}\pm 2q_{y}-2\right) +T \\
B_{y}\approx \mp 4tq_{y}+i\gamma%
\end{array}%
\right. ,
\end{equation}%
and the dispersion along the $k_{x}=0$ ($k_{y}=0$) direction is square-root
(linear) as%
\begin{eqnarray}
E_{+}\left( k_{x}=0\right) &\approx &2\sqrt{\mp JT\mp 2J^{2}\mp 2i\gamma t}%
\sqrt{q_{y}}, \\
E_{+}\left( k_{y}=0\right) &\approx &\sqrt{-2JT-4J^{2}}\left\vert
q_{x}\right\vert .
\end{eqnarray}%
For $k_{\mathrm{c}x}=\pi $ (i.e., $T\pm \gamma =2J$), we approximately have $%
\mathbf{B}(\mathbf{k})$ field near $\left( \pi ,\pm \pi /2\right) $ as%
\begin{equation}
\left\{
\begin{array}{l}
B_{x}\approx J\left( q_{x}^{2}\mp 2q_{y}-2\right) +T \\
B_{y}\approx \pm 4tq_{y}+i\gamma%
\end{array}%
\right. ,
\end{equation}%
and the dispersion along the $k_{x}=0$ ($k_{y}=0$) direction is square-root
(linear) as%
\begin{eqnarray}
E_{+}\left( k_{x}=0\right) &\approx &2\sqrt{\mp JT\pm 2J^{2}\pm 2i\gamma t}%
\sqrt{q_{y}}, \\
E_{+}\left( k_{y}=0\right) &\approx &\sqrt{2JT-4J^{2}}\left\vert
q_{x}\right\vert .
\end{eqnarray}

Thirdly, the system is Hermitian at $\gamma =0$ and the BTPs are DPs. The
dispersion near the DPs with $w_{\mathrm{I}}=0$ is anisotropy, being linear
along one direction and quadratic along the perpendicular direction. DPs
here are semi-Dirac points \cite{QChen}. For $k_{\mathrm{c}x}=\pm \pi /2$
(i.e., $T=0$), we approximately have $\mathbf{B}(\mathbf{k})$ field near $%
\left( \lambda _{1}\pi /2,\lambda _{2}\pi /2\right) $ as%
\begin{equation}
\left\{
\begin{array}{l}
B_{x}\approx -2J\left( \lambda _{1}q_{x}+\lambda _{2}q_{y}\right) \\
B_{y}\approx 4\lambda _{1}\lambda _{2}tq_{x}q_{y}%
\end{array}%
\right. ,
\end{equation}%
and the dispersion along the $\lambda _{1}k_{x}=\lambda _{2}k_{y}$ ($\lambda
_{1}k_{x}=-\lambda _{2}k_{y}$) direction is linear (quadratic) as%
\begin{eqnarray}
E_{+}\left( \lambda _{1}k_{x}=\lambda _{2}k_{y}\right) &\approx
&4J\left\vert q_{x}\right\vert , \\
E_{+}\left( \lambda _{1}k_{x}=-\lambda _{2}k_{y}\right) &\approx
&4tq_{x}^{2}.
\end{eqnarray}%
\ For $k_{\mathrm{c}x}=0$ (i.e., $T=-2J$), we approximately have $\mathbf{B}(%
\mathbf{k})$ field near $\left( 0,\pm \pi /2\right) $ as%
\begin{equation}
\left\{
\begin{array}{l}
B_{x}\approx -J\left( q_{x}^{2}\pm 2q_{y}\right) \\
B_{y}\approx \mp 4tq_{y}%
\end{array}%
\right. ,
\end{equation}%
and the dispersion along the $k_{x}=0$ ($k_{y}=0$) direction is linear
(quadratic) as%
\begin{eqnarray}
E_{+}\left( k_{x}=0\right) &\approx &2\sqrt{J^{2}+4t^{2}}\left\vert
q_{y}\right\vert , \\
E_{+}\left( k_{y}=0\right) &\approx &Jq_{x}^{2}.
\end{eqnarray}%
For $k_{\mathrm{c}x}=\pi $ (i.e., $T=2J$), we approximately have $\mathbf{B}(%
\mathbf{k})$ field near $\left( \pi ,\pm \pi /2\right) $ as%
\begin{equation}
\left\{
\begin{array}{l}
B_{x}\approx J\left( q_{x}^{2}\mp 2q_{y}\right) \\
B_{y}\approx \pm 4tq_{y}%
\end{array}%
\right. ,
\end{equation}%
and the dispersion along the $k_{x}=0$ ($k_{y}=0$) direction is linear
(quadratic) as%
\begin{eqnarray}
E_{+}\left( k_{x}=0\right) &\approx &2\sqrt{J^{2}+4t^{2}}\left\vert
q_{y}\right\vert , \\
E_{+}\left( k_{y}=0\right) &\approx &Jq_{x}^{2}.
\end{eqnarray}

Fourthly, we consider the dispersion near DPs with $w_{\mathrm{I}}=\pm 1$.
Near $\left( k_{\mathrm{c}x},\pm \pi /2\right) $ with $k_{\mathrm{c}x}\neq 0$%
, $\pi $, $\pm \pi /2$, we approximately have $\mathbf{B}(\mathbf{k})$ field
as
\begin{equation}
\left\{
\begin{array}{l}
B_{x}\approx 2J\left( -q_{x}\sin k_{\mathrm{c}x}\mp q_{y}\right) \\
B_{y}\approx \mp 4tq_{y}\cos k_{\mathrm{c}x}%
\end{array}%
\right. ,
\end{equation}%
and the dispersion along the $k_{y}=\xi k_{x}$ ($k_{x}=-\xi k_{y}$)
direction is%
\begin{equation}
\begin{array}{c}
E_{+}\left( k_{y}=\xi k_{x}\right) \approx 2\sqrt{J^{2}\left( \sin k_{%
\mathrm{c}x}\pm \xi \right) ^{2}+4t^{2}\xi ^{2}\cos ^{2}k_{\mathrm{c}x}}%
\left\vert q_{x}\right\vert \\
E_{+}\left( k_{x}=-\xi k_{y}\right) \approx 2\sqrt{J^{2}\left( \xi \sin k_{%
\mathrm{c}x}\mp 1\right) ^{2}+4t^{2}\cos ^{2}k_{\mathrm{c}x}}\left\vert
q_{y}\right\vert%
\end{array}%
.
\end{equation}%
Notably, DPs with $w_{\mathrm{I}}=\pm 1$ have linear dispersion along any
direction; DPs here are Dirac points.


\begin{thebibliography}{999}
\bibitem{Note} The topological band structure determines the EP chirality,
which is relevant to the coalescence eigenstate. In Fig. \ref{fig4}(b), $w_{%
\mathrm{II}}$\ from left to right are $+\left( -\right) \left\{
-1/2,0,+1/2\right\} $\ when $\gamma >0$\ ($\gamma <0)$; however, the two
outer EPs have the identical coalescence state.
\end{thebibliography}
\end{document}